\documentstyle[12pt]{article}
\newcommand{\be}{\begin{equation}}
\newcommand{\ee}{\end{equation}}
\newcommand{\bea}{\begin{eqnarray}}
\newcommand{\eea}{\end{eqnarray}}

\newcommand{\Gm}{\Gamma}
\newcommand{\gm}{\gamma}

\newcommand{\eps}{\epsilon}

\newcommand{\dd}{\mbox{d}}
\newcommand{\nn}{\nonumber}
\begin{document}
\vspace{1cm}
\begin{center}
\Large CALCULATION OF FEYNMAN DIAGRAMS WITH ZERO MASS
THRESHOLD FROM THEIR SMALL MOMENTUM EXPANSION \\

\vspace{0.9cm}

\large J.~FLEISCHER$^a$\footnote{~E-mail:
fleischer@physik.uni-bielefeld.de},
~~V.A.~SMIRNOV$^b$\footnote{~Supported by European project
Human Capital and Mobility under CHRX-CT94-0579 \\ ~E-mail:
smirnov@theory.npi.msu.su}~~and~~O.V.~TARASOV$^a$\footnote
{~Supported by Bundesministerium f\"ur Forschung und Technologie
under PH/05-6BI93P.
~On leave of absence from Joint Institute for Nuclear Research,
141980 Dubna, Moscow Region, Russian Federation.
Present address: IfH~ DESY, Zeuthen,Platanenallee 6,
15738, Germany,E-mail: tarasov@ifh.de.} \\
\baselineskip=13pt
\normalsize $^a$Fakult\"at f\"ur Physik, Universit\"at Bielefeld \\
\baselineskip=12pt
D-33615 Bielefeld , Germany \\
\baselineskip=13pt
$^b$Nuclear Physics Institute of Moscow State University \\
\baselineskip=12pt
119899 Moscow, Russian Federation
\end{center}

\abstract{A method of calculating Feynman diagrams from their
small momentum expansion \cite{ft} is extended to diagrams with zero
mass thresholds.
We start from the asymptotic expansion in large masses \cite{ae}
(applied to the case when all $M_i^2$ are large compared to all 
momenta squared). Using dimensional regularization,
a finite result is obtained in terms of powers of logarithms (describing
the zero-threshold singularity) times power series in the momentum
squared. Surprisingly, these latter ones represent functions, which not
only have the expected physical
``second threshold'' but have a branchcut singularity
as well below threshold at a mirror position.
These can be understood as pseudothresholds
corresponding to solutions of the Landau equations.
In the spacelike region the
imaginary parts from the various contributions cancel.
For the two-loop examples with one mass $M$, in  the timelike region
for $q^2 \approx M^2$ we obtain approximations of high precision. This will
be of relevance in particular for the calculation of the decay
$Z \to b\bar{b}$  in the $m_b=0$ approximation.}

\newpage

\section{Introduction}

   Once it has been observed \cite{ft} that the calculation of Feynman diagrams
on their cut can be performed with high precision from their Taylor expansion
coefficients, there are several advantages of this method, which make it
really quite attractive: firstly, the Taylor coefficients being known,
the remaining calculation of the diagram in the whole complex plane is a
relatively easy task. Secondly, more important, the precision with which the
coefficients can be calculated (from vacuum diagrams) is practically unlimited
(e.g. 50 to 100 decimals with the multiple precision program of \cite{Bail}
is ``standard''). This last property is of particular relevance in higher
loop orders when many diagrams
(of the order of 1000) contribute, namely the high precision of the Taylor
coefficients suggests that in such a case the scalar amplitudes should be
added on the level of their Taylor coefficients. Finally, as mentioned in
\cite{ft}, in the 2-loop 3-point case, in which we are mainly interested here,
there occur in general 10 numerator scalar products of 4 momenta, but only 9
(internal or external) lines against which to cancel these. This causes
serious problems in the evaluation of two-loop vertex Feynman diagrams which
are not present if only bubble integrals are to be evaluated like in the 
Taylor expansion.
For these reasons it appears worthwhile to develop and extend the method of 
Taylor expansion further to make it applicable to the various kinematical 
situations in the Standard Model.

   The purpose of the present paper is to demonstrate an extension of the
previous method \cite{ft}, which can be applied for vertex diagrams with massless
thresholds. In such a case the Taylor expansion  of the Feynman diagrams
does not exist because of logarithmic singularities at zero momenta
squared.
The method which we propose for this type of diagrams is a combination of
using standard explicit formulae for asymptotic 
expansions in large masses \cite{ae} (see \cite{vs} for a short review) 
and the summation procedure of \cite{ft}. Thus, in the large mass limit
($M^2 \gg \mid q^2 \mid$)~ we get power ~series in ~$q^2/M^2$~ factorizing powers
of \\
ln(-$q^2/M^2$). These power series can be summed by means of  Pad\'{e}
approximants such that the validity of this expansion is extended to large $q^2/M^2$,
in the spacelike region as well as in the timelike region (due to
conformal mapping). Note that the general formulae for
asymptotic expansions in momenta and masses \cite{ae}
have been successfully applied in a number of papers
\cite{appl,dst,bdst}. In particular, two-loop self-energy diagrams
with general masses were
calculated in \cite{dst}
in the region of small and large momentum, and in \cite{bdst}
in the case of massless thresholds and thresholds with small masses.

   Our paper is organized as follows: in Sect.~2 an introductory example of
a ~two-loop ~self-energy diagram ~is ~given ~for ~which ~the ~factorization ~of ~the \\
ln(-$q^2/M^2$) is explicitly known. In Sect. 3 we show cases of interest
for the decay $Z \to b\bar{b}$ and select typical examples for the 
demonstration of our method. Sect.~4 recalls the general method and 
demonstrates the calculation of the ``naive'' part. In Sects. 5 and 6 
the above examples are worked out explicitly and Sect. 7 contains our 
conclusions.
   
\section{An example of a two-loop self-energy \newline diagram
with zero threshold}

Before turning to complicated vertex diagrams we demonstrate
how the summation by Pad\'{e} approximants works
in a simpler case of the self-energy integral
$\widetilde{I}_3$ (see Fig.~1) with zero threshold \cite{bft}, 
where an explicit result written in the form of the large
mass expansion is known:
\begin{equation}
 \widetilde{I}_3=\widetilde{f}_0\left( \frac{q^2}{M^2} \right)
 +\widetilde{f}_1 \left( \frac{q^2}{M^2} \right)
  \ln \left( - \frac{q^2}{M^2} \right)
\end{equation}
with

\begin{eqnarray}
&&\widetilde{f}_0\left( \frac{q^2}{M^2} \right)
 =\frac{3}{2M^2} \sum_{n=0}^{\infty}
 \frac{n! \Gamma(\frac32)}
 {\Gamma(n+\frac32) (n+1)^2} \left( -\frac{q^2}{4M^2} \right)^n \\
&& \nonumber\\
&&\widetilde{f}_1\left( \frac{q^2}{M^2} \right)
 =-\frac{1}{2M^2} \sum_{n=0}^{\infty}
 \frac{n! \Gamma(\frac32)}
 { \Gamma(n+\frac32) (n+1)}\left( -\frac{q^2}{4M^2} \right)^n .
\end{eqnarray}

 Summing the series for $\widetilde{f}_0$ and $\widetilde{f}_1$
 in the ``standard'' manner \cite{ft}, the results of Table 1.
are obtained.
They  demonstrate that high precision numerical values can be
obtained for a large range of $q^2$ values in this way.

\section{Two-loop vertex diagrams with zero \newline thresholds}

 Concerning the vertex diagrams, there are many different topologies
contributing to a 3-point function in the Standard Model. For our
purpose of demonstrating the method, we confine ourselves to the 
``planar'' case with the topology shown in Fig. 2 (see ``generic'').
Distributing in all possible ways
massive particles over the six virtual lines of this figure, one finds 
22 diagrams with a zero mass threshold. Our main interest
in the present study is, however, besides
the development of a new method, its possible application to the calculation
of the decay amplitude for $Z \to b\bar{b}$. For this process $m_b = 0$ can be
considered as a good approximation and
for the corresponding kinematical situation
($p_1^2=p_2^2=0$) the new approach is
directly applicable  (see also \cite{ft}). Therefore,
instead of presenting all possible 22 diagrams with zero mass thresholds, in
Figs. 2 and 3 we give only those 
contributing to $Z \to b\bar{b}$ (with $m_b = 0$ and $m_t$ large, neglecting
quark mixing).
Fig.~3 presents infrared divergent diagrams and they are merely given 
for completeness:  $Cases$ 7,8 and 9 will be considered in a separate
publication.
$Case$ 10 is a diagram with massless particles only, and
our method is of
no relevance here. This diagram (and also other topologies), however,
has been evaluated in \cite{Gons} (see also \cite{KraLa}).
In Figs. 2 and 3 we choose all non-zero masses to be equal ($=M$).

   Typical examples which we work out here explicitly are $Case$ 1 and
$Case$ 5 of Fig.~2. These have in the expansion coefficients terms of 
the order $1/{\varepsilon}$ ($Case$ 1) and $1/{\varepsilon}$ and 
$1/{\varepsilon}^2$ ($Case$ 5), respectively
$(d=4-2\varepsilon)$. Taking into account factors of the form
$ \left( \frac{{\mu}^2}{M^2} \right)^{\eps} $ and 
$ \left( -\frac{{\mu}^2}{q^2} \right)^{\eps}$ ($q=p_1-p_2$),
(see below) and expanding
these in terms of $\varepsilon$, the poles in $\varepsilon$ as well as the 
dependence on the scale parameter $\mu$ drop out. In the final result
there remain terms factorizing $ln(-q^2/M^2)$ and  $ln^2(-q^2/M^2)$
from the latter expansion after cancellation of the
corresponding powers of $\varepsilon$.

\section{Large mass expansion and calculation of the ``naive'' part}

   The large mass limit is obtained in the following manner: if some masses
are much greater than the other masses (in our case all the small masses are
zero) and all the momenta, one has \cite{ae}

\begin{equation}
F_{\Gamma} (p_1,p_2, M; \epsilon)
\; \stackrel{\mbox{\footnotesize$M \to \infty$}}{\mbox{\Large$\sim$}} \;
\sum_{\gamma} F_{\Gamma / \gamma} (p_1,p_2;\epsilon)
\circ {\cal T}_{q^{\gamma}}
F_{\gamma} (q^{\gamma}, M;\epsilon),
\label{LME}
\end{equation}
$F_{\ldots}$ standing for (sub-)diagrams and reduced diagrams
characterized by their index: $\Gamma$
for the original diagram, $\gamma$ a subdiagram ($\gamma \subset \Gamma$)
and $\Gamma / \gamma$ obtained from the original diagram by factorizing the
product of scalar propagators as
$\Pi_{\Gamma} \equiv \Pi_{\Gamma / \gamma} \Pi_{\gamma}$ such that more
explicitly we have ($l$ is the number of loops)

\begin{equation}
F_{\Gamma / \gamma} \circ {\cal T}_{q^{\gamma}} F_{\gamma} =
\int dk_{1} \cdots  dk_{l} \Pi_{\Gamma / \gamma} {\cal T}_{q^{\gamma}} \Pi_{\gamma}.
\label{Scp}
\end{equation}

Here ${\cal T}_{\ldots}$ is the operator of the Taylor expansion
w. r. t. the set of external momenta $q^{\gamma}$ of the subgraph $\gamma$.
The summation in (\ref{LME}) is performed over the following subgraphs:\\

   ---~~~each $\gamma$ (it may be disconnected) contains all the lines with
large masses,\\

   ---~~~each $\gamma$ is 1PI w. r. t. lines with small masses.\\

   The (``naive'') contribution from the original diagram $\Gamma$ itself
is obviously always contained
in the $\sum_{\gamma}$. In various terms of this sum the integrations
of the type (\ref{Scp}) yield in general divergent coefficients of the
asymptotic expansion.
These divergences are both of infrared
and ultraviolet nature, the latter being due
to high powers of integration momenta produced by the
${\cal T}_{\ldots}$ operator. Therefore (\ref{LME}) is to be understood
in terms of some regularization for which we take
dimensional regularization. This is so
even if the original diagram is convergent: summing all contributions
($\sum_{\gamma}$) the divergent terms and those depending on the scale
parameter $\mu$ (from dimensional regularization) must cancel. This will
be used below as a strong check of our calculational procedure.

   In the following we will calculate the diagrams $Case$ 1 and $Case$ 5
of Fig.~2. These are typical in the sense that they have one and two zero mass
thresholds, resulting in the above mentioned
terms up to $1/ \varepsilon$ and $1/ {\varepsilon}^2$
in the Taylor coefficients, respectively or $ln(-q^2/M^2)$ and
$ln^2(-q^2/M^2)$ in the final result.
The three particle zero mass threshold in $Case$ 6 does not
induce higher than $1/ {\varepsilon}^2$ terms in the Taylor coefficients either.

   The ``higher'' terms in the $\sum_{\gamma}$ (i.e. all except for the naive
one) can be handled in a straightforward though tedious calculation. The reason
for the relative simplicity is that only factorizing massive one-loop bubble 
(vacuum) integrals and massless propagator type integrals occur. Results for 
these higher terms  will be given in the following sections.

   The situation is different, however, for the ``naive'' contribution. 
In this case the approach of \cite{ZiF} turns out to be particularly adequate.
The general expansion of (any loop) scalar 3-point function with its
momentum space representation $C(p_1, p_2)$ can be written as 
\begin{equation}
\label{eq:exptri}
C(p_1, p_2) = \sum^\infty_{l,m,n=0} a_{lmn} (p^2_1)^l (p^2_2)^m
(p_1 p_2)^n ,
\end{equation}
where the coefficients  $a_{lmn}$ are to be determined from the given diagram.
As we will show, in the cases under consideration the representation of the
Taylor coefficients given in \cite{ZiF} yields
only ``genuine two-loop bubbles'' but no factorizing one-loop ones. Introducing
the abbreviations (see also Fig.~2)
$c_1=k_1^2-m^2_1, c_2=k_1^2-m^2_2,
c_3=k_2^2-m^2_3, c_4=k_2^2-m^2_4$ and $c_5=k_2^2-m^2_5, c_6=(k_1-k_2)^2-m^2_6$,
we have for the expansion coefficients

\begin{equation}
\label{e24}
(i\pi^2)^2 a_{00n} = \frac{2^n}{n+1} ({\mu}^2)^{2 \varepsilon}
\int \dd^d k_1 \dd^d k_2 F_n \cdot
\frac{1}{c_1~ c_2~ c_3 ~ c_4~ c_5 ~ c_6},
\end{equation}
where $\mu$ is the scale parameter of dimensional regularization.

In general the vertex diagram depends on
three external momenta squared (see \ref{eq:exptri}), 
each of which is an independent expansion variable.
Putting $p_1^2 = p_2^2 = 0$ the corresponding summation indices are also zero.
In (\ref{e24}) $F_n$ is given by

\bea
F_n
=
\sum_{\nu,\nu^\prime,\mu^\prime} a^{n\mu^\prime}_{\nu\nu^\prime}
\frac{(k^2_1)^{n-(\nu + \nu^\prime)+\mu^\prime}}
     {c_1^{n-\nu} c_2^{n-\nu^\prime}           }
\frac{(k^2_2)^{\mu^\prime}    }
     {c_3^\nu c_4^{\nu^\prime}}
\frac{1}
     {2^{\nu+\nu^\prime -2{\mu^\prime}}} 
\left\{( k^2_1 + k^2_2   -m^2_6
)^{\nu + \nu^\prime -2{\mu^\prime}} \right. \nonumber \\ \left.
- \sum_{\alpha = 1,odd}^{\nu + \nu^\prime -2{\mu^\prime}}
(k_1^2 + k^2_2 - m^2_6)^{\nu + \nu^\prime - 2{\mu^\prime} - \alpha}
(2 k_1  k_2)^{\alpha-1} \cdot c_6\right\} ,
\eea
the coefficients $ a^{n{\mu^\prime}}_{\nu\nu^\prime} $ being known explicitly for
arbitrary $d$ \cite{ZiF}.
We see that due to cancellation of $c_6$ in the above sum over
$ \alpha $ this contribution contains only factorizing one-loop terms which vanish
for the above mass combinations in dimensional regularization. Thus it
remains to calculate the "genuine two-loop" contributions.

In Case 1 ($m_6 = M$) we write $(\lambda = \nu + \nu^\prime - 2{\mu^\prime} )$

\begin{eqnarray*}
(k^2_2)^{{\mu^\prime}} (k^2_1 + k^2_2 - M^2)^{\lambda}
 = \sum^{\lambda}_{\beta = 0} \sum^{{\mu^\prime}}_{\gamma =0}
{\lambda \choose \beta} {{\mu^\prime} \choose \gamma }
(M^2)^{{\mu^\prime} - \gamma} (k^2_1)^{\lambda - \beta} c_3^{\beta + \gamma}
\end{eqnarray*}
and obtain

\bea
a_{00n} = \frac{2^n}{n+1} \sum_{\nu,\nu^\prime,{\mu^\prime}} a^{n{\mu^\prime}}_{\nu\nu^\prime}
\frac{1}{2^{\lambda}}
\sum^{\lambda}_{\beta = 0} \sum^{{\mu^\prime}}_{\gamma =0}
{\lambda \choose \beta} {{\mu^\prime} \choose \gamma }
\nn \\
\times
(M^2)^{{\mu^\prime} - \gamma}
\frac{({{\mu}^2})^{2 \varepsilon}}{(i\pi^2)^2}\int \dd^d k_1 \dd^d k_2
\frac{1}{(k^2_1)^{{\nu}_1} {c_3}^{{\nu}_2} c_6}
\label{exy}
\eea
with ${\nu}_1 = n - (\nu+\nu^\prime) + {\mu^\prime} +\beta +2$ and
${\nu}_2 = (\nu+\nu^\prime) - \beta - \gamma +3$. The two-loop
bubble integrals can be evaluated explicitly:

\begin{eqnarray*}
& &\frac{1}{(i\pi^2)^2}\int \dd^d k_1 \dd^d k_2
\frac{1}{(k^2_1)^{{\nu}_1} {c_3}^{{\nu}_2} c_6}\\
& &=(-1)^{(\nu_1 + \nu_2 +1)}
\frac{\Gamma({\nu}_1 + {\nu}_2 +1 -d) \Gamma(\frac{d}{2} - {\nu}_1)
      \Gamma({\nu}_1 + {\nu}_2 - \frac{d}{2}) \Gamma({\nu}_1 +1 - \frac{d}{2})}
     {\Gamma({\nu}_2) \Gamma(\frac{d}{2}) \Gamma(2 {\nu}_1 + {\nu}_2 +1 -d)
      (M^2)^{({\nu}_1 + {\nu}_2 +1 -d)}}.
\end{eqnarray*}
Expanding in $\varepsilon$, a divergent $\frac{1}{\varepsilon}$ - term is
obtained, which in this case comes from the infrared divergence of the
above integral.\\

Similarly as above we calculate the "naive" part for Case 5: with

\begin{eqnarray*}
(k^2_2)^{{\mu^\prime}} (k^2_1 + k^2_2 - M^2)^{\lambda}
 = \sum^{\lambda}_{\beta = 0} \sum^{\beta}_{\gamma =0}
{\lambda \choose \beta} {\beta \choose \gamma }
(-M^2)^{\beta - \gamma} (k^2_1)^{\lambda - \beta} (k^2_2)^{{\mu^\prime} + \gamma}
\end{eqnarray*}
we have
\bea
a_{00n} = \frac{2^n}{n+1} \sum_{\nu,\nu^\prime,{\mu^\prime}} a^{n{\mu^\prime}}_{\nu\nu^\prime}
\frac{1}{2^{\lambda}}
\sum^{\lambda}_{\beta = 0} \sum^{\beta}_{\gamma =0}
{\lambda \choose \beta} {\beta \choose \gamma }
\nn \\
\times
(-M^2)^{\beta - \gamma}
\frac{({{\mu}^2})^{2 \varepsilon}}{(i\pi^2)^2}\int \dd^d k_1 \dd^d k_2
\frac{1}{(k^2_1)^{{\nu}_1} {c_3}^{(\nu_2-1)} c_5 c_6}
\label{eyz}
\eea
with ${\nu}_1$ and ${\nu}_2$ as above. The situation is now somewhat
more complicated due to the fact that $m_3 \neq m_5$ and the following
partial fraction decomposition needs to be performed ( $p={\nu}_2-2$,
$m_3 = 0$ and $m_5 = M)$:

\[
\frac{1}{c_3^{p+1} c_5} = - \sum^p_{i=0} \frac{1}{(M^2)^{p+1-i}}
\frac{1}{c_3^{i+1}} + \frac{1}{(M^2)^{p+1}} \frac{1}{c_5}\ ,
\]
yielding the two-loop bubble Integrals

\begin{eqnarray*}
& &\frac{1}{(i\pi^2)^2}\int \dd^d k_1 \dd^d k_2
\frac{1}{(k^2_1)^{{\nu}_1} {c_3}^{i+1} c_6}=\\
& &(-1)^{(\nu_1  + i)}
\frac{\Gamma({\nu}_1 + i +2 -d) \Gamma(\frac{d}{2} - i - 1)
      \Gamma({\nu}_1 + i + 1 - \frac{d}{2}) \Gamma(\frac{d}{2} - {\nu}_1) }
     {\Gamma({\nu}_1) \Gamma(\frac{d}{2}) \Gamma(i+1)
      (M^2)^{(\nu_1 + i +2 -d)} }
\end{eqnarray*}
and

\begin{eqnarray*}
& &\frac{1}{(i\pi^2)^2}\int \dd^d k_1 \dd^d k_2
\frac{1}{(k^2_1)^{{\nu}_1} c_5 c_6}=\\
& &(-1)^{\nu_1}
\frac{\Gamma({\nu}_1  +2 -d) \Gamma(\frac{d}{2} - {\nu}_1)
      \Gamma^2({\nu}_1 + 1 - \frac{d}{2}) }
     {\Gamma(\frac{d}{2}) \Gamma(2 {\nu}_1 + 2  -d)
      (M^2)^{(\nu_1  +2 -d)} }.
\end{eqnarray*}
Expanding in $\varepsilon$, in this case $\frac{1}{\varepsilon^2}$-terms occur.

\section{Threshold singularity of the 
type \newline $\ln (-q^2/M^2)$ $(Case\; 1)$}

   According to Fig.~2 there is only one subdiagram $\gamma$ which contains
all four massive lines, namely the box. Accordingly the only contribution
with $\gamma \not= \Gamma$ reads

\bea
(\mu^2)^{2\varepsilon}
\int \frac{\dd^d k_1}{k_1^2 (q-k_1)^2 }
\int \frac{\dd^d k_2}{(k_2^2-M^2)} \times \nn \\
\times {\cal T}_{p_1,q,k_1} \frac{1}{[(k_1-k_2)^2-M^2][(q+k_2)^2-M^2] [(p_1-k_2)^2-M^2]} .
\label{LME1}
\eea

   Here the external momenta for the box subgraph 
   are besides $p{_1}$ and $q$
also the loop momentum $k_1$. Note that, for the contributions
of each subgraph
$\gm$ in (\ref{LME}) one can choose loop momenta in an appropriate way.
For example, in (\ref{LME1}) the rooting of the loop momenta is
different from the ``naive'' contribution
(observe also below similar rerooting in $Case$ 5).

The evaluation of (\ref{LME1}) results in products
of one-loop massive bubble integrals with monomials in the numerator and one-loop
propagator-type massless integrals, the latter yielding the factor
$(-1/q^2)^{\varepsilon}$. Explicitly we obtain

\be
\frac{1}{(4\pi)^d} \frac{1}{M^4} \left( \frac{{\mu}^2}{M^2} \right)^{\eps}
\left( -\frac{{\mu}^2}{q^2} \right)^{\eps}
\sum_{n=0}^{\infty} c^{(2)}_n (\eps) (q^2/M^2)^n ,
\label{C20}
\ee


where
\bea
c^{(2)}_n (\eps) =
\sum_{{i_1,i_2,n_3 \geq 0,\; i_1+i_2+n_3 \; \mbox{\scriptsize even}}
\atop {i_1+i_2+n_3 \leq 2n}} \sum_{j_3 \geq 0}^{(n_3,-2)}
(-1)^{(i_1+i_2+n_3)/2} \;\frac{(n-(i_1+i_2-n_3)/2)!}{(n-(i_1+i_2+n_3)/2)!}
\nn \\
\times \frac{i_1!i_2! \theta (i_1+i_2-j_3)
\theta (i_1-i_2+j_3) \theta (-i_1+i_2+j_3)}{((n_3-j_3)/2)!
((i_1+i_2-j_3)/2)! ((i_1-i_2+j_3)/2)! ((-i_1+i_2+j_3)/2)!}
\nn \\
\times
\frac{\Gm(\eps)\Gm(1-\eps)
\Gm((i_1+i_2-j_3)/2+1-\eps)}{\Gm((i_1+i_2-j_3)/2+2-2\eps)}
\nn \\
\times
C(n+(i_1+i_2+n_3)/2+4; (i_1+i_2+n_3)/2)
\label{nonn}
\eea
and
\be
C(r,s) = 
\Gm\left( r-s-\frac{d}{2} \right) / \Gm(r) .
\label{CRS}
\ee

   The $\varepsilon-$poles in this case come from ultraviolet divergences.
Adding the naive and the above contribution yields (up to a factor
$(1/{16 {\pi}^2})^2$)

\begin{eqnarray}
F_{\Gamma} (q^2,M^2) = \frac{1}{M^4} \left\{
\sum_{n=0}^{\infty} f_{0n} (q^2/M^2)^n +
\sum_{n=0}^{\infty} f_{1n} (q^2/M^2)^n \ln (-q^2/M^2) \right\} \nn \\
\label{AE2}
\equiv \frac{1}{M^4} \left\{f_0 (q^2/M^2 )
+ f_1 (q^2/M^2 ) \ln (-q^2/M^2) \right\},
\end{eqnarray}
i.e. all $\varepsilon-$poles and the scale parameter ${\mu}$
have cancelled, which at the same time serves as a  helpful check of the
correctness of the calculation.
Evaluating the complicated coefficients and summing all
contributions was performed with FORM \cite{FORM}.
   For the expansion coefficients of  $f_1$  we found
by inspection the following recurrence relation:

\begin{eqnarray}
&& 8(2n+5)(2n+7)(2n+9)(n+4)~f_{1,n+3}= \nonumber \\
&& \nonumber \\
&& ~~~~~~-4(2n+5)(2n+7)(n+4)^2 ~f_{1,n+2}  \nonumber \\
&& \nonumber \\
&&~~~~~~  +2(2n+5)(n+3)(n+2)^2~f_{1,n+1}+(n+3)(n+2)^3~ f_{1,n}~~,
\end{eqnarray}
from which we obtain:
\begin{equation}
f_{1,n}=\frac{  \Gamma(n+2) \Gamma(\frac52) }
    {6 (-4)^ n\Gamma(n+\frac52) } \sum_{j=0}^{n} \frac{1+2 (-1)^j}{(j+1)^2}.
\end{equation}

This explicit form allows to sum the series for $f_1$,
yielding the following integral representation:

\be
f_1=\sum_{n=0}^{\infty} \left( -\frac{q^2}{M^2} \right)^n~f_{1,n}
 =\frac{M^2}{q^2} \int_0^1 \ln z dz
 \left[\phi\left( z,\frac{q^2}{M^2}\right)
-2 \phi\left( -z,\frac{q^2}{M^2}\right)\right]
\label{closef}
\ee
where
\begin{equation}
\phi(z,x)=\frac{4 \arcsin( \sqrt{zx/4})}{(1-z) \sqrt{zx(4+zx)} } .
\end{equation}

An interesting feature of this representation
is that two thresholds,  one at $q^2=4M^2$
and another at $q^2=-4M^2$ can immediately be read off.
There ought to be, however, only one threshold at $q^2=4M^2$.
Indeed it turns out that $f_0$ has a singularity
at $q^2=-4M^2$ as well and that for  $q^2 < -4M^2$ the
imaginary part of $f_1 \ln(-q^2/M^2)$
and $f_0$  cancel.

Note that the position of the ``mirror'' threshold of the functions
$f_0$ and $f_1$ exactly corresponds to a
pseudothreshold of the given Feynman diagram which
is  a solution of the Landau equations ( see e.g. ~\cite{LaLi}).
In fact solving these, we obtain for $m_1=m_2=0$ the pseudothreshold
$q^2=-(m_5+m_6)^2$. 
\footnote{We are grateful to J.B.~Tausk
who has drawn our attention to this property and provided
several examples, including the following one.}
An even simpler case is the corresponding 
one-loop vertex with a massive line connecting two massless ones. 
This graph is proportional to ( x=$q^2/M^2$ )

\begin{equation}
  Li_2(1+x)-\zeta(2) = -Li_2(-x)-\ln(1+x) \ln(-x),
\end{equation}
where in the second form of writing the logarithmic singularity at
x=0 has been isolated and now again the pseudothreshold singularity
at x=-1 appears in the separate terms. For two-loop self-energy 
diagrams one obtains by inspection corresponding results from Refs.
~\cite{bdst} and ~\cite{st}. At this point it is interesting to note
that the self-energy diagram $\widetilde{I}_3$, dealt with in Sect.2,
has no pseudothreshold, i.e. no such solution of the Landau equations 
exists.

We have found, however, a closed form as in (\ref{closef}) only for $f_1$
and not for $f_0$. Therefore, to demonstrate the above statement, we have
to rely on our numerical approach. Results for $Case$ 1 are given in Tables
2 to 4 (in all tables the results are
given up to a factor $\frac{1}{(16{\pi}^2)^2 M^4}$). 

Table 2
gives results for spacelike $q^2$. For $0 > q^2 \ge -4 M^2$
we achieve fast convergence, calculating the indicated  Pad\'{e} approximants,
taking into account Taylor coefficients up to n=30. Of course the
integral is real in this domain and agrees excellently with the Monte Carlo
control calculation (five dimensional integration over Feynman parameters
with an estimated relative error $< {10}^{-4}$).

   For $q^2 < -4M^2$ we first performed a conformal mapping in terms of
an ``$\omega$-transform'' \cite{ft} before using Pad\'{e} approximants
to sum the series for $f_0$ and $f_1$. This is of course always necessary
if one wants to calculate a function on a cut.
As a result we obtain in the above domain small imaginary parts, which are
due to insufficient cancellation of the two imaginary parts. It is seen,
however, that these become quickly small with increasing
order of the approximants. The situation becomes even more amazing if
one looks at the imaginary parts of $f_0$ and $f_1$ separately: they do
increase with $q^2$ relative to their real parts as is demonstrated in
Table 4! Nevertheless, in this manner we obtain even for large negative
values like $q^2/M^2=-50$ an accuracy of at least 3 decimals as is also
confirmed by the MC control calculation.


   Table 3 gives our results in the timelike region. The analytic continuation
of the logarithm requires in this case to write
\begin{eqnarray}
ln(-q^2/M^2)= ln\mid q^2/M^2 \mid - i \pi
\label{ln}
\end{eqnarray}
so that an imaginary part is obtained for $q^2 > 0$. Table 3 gives only
results above the second threshold since below the convergence is even better:
at $q^2 = M^2$ we obtain a precision of 9 decimals with merely 9 coefficients
([4/4]). A precision of at least 3 decimals is achieved both for the real
and imaginary part up to $q^2/M^2$=50. This can be concluded from the
convergence properties of the approximants.

\section{Threshold singularity of the 
type \newline $ln^2(-q^2/M^2)$ $(Case\; 5)$}

There are two massless cuts
so that we shall have the double logarithm in the expansion.
The set of subgraphs in this case is given by
$\gm_1=\Gm$ and the higher terms with  $\gamma \ne \Gamma$ : $\gm_2=\{3456\},
\gm_3=\{1256\}$, and $\gm_4=\{56\}$. Note that $\gm_3$ and $\gm_4$ are
disconnected. The subgraph $\gm_1$ was discussed in Sect. 4. 
In terms of integrals of the form (\ref{Scp}) we have

for $\gm_2$
\bea 
(\mu^2)^{2\varepsilon}
\int \dd^d k_1 \frac{1}{k_1^2 (q-k_1)^2} \nn \\ 
\times \int \dd^d k_2 \frac{1}{k_2^2}
{\cal T}_{p_1,q,k_1} \frac{1}{((p_1+k_2)^2-M^2)((k_1-k_2)^2-M^2)(q-k_2)^2},
\eea

for $\gm_3$
\bea 
(\mu^2)^{2\varepsilon}
\int \dd^d k_2 \frac{1}{k_2^2 (q-k_2)^2}
{\cal T}_{p_1,k_2} \frac{1}{(p_1+k_2)^2-M^2}
\nn \\ \times
\int \dd^d k_1 \frac{1}{k_1^2}
{\cal T}_{q,k_2} \frac{1}{((k_1-k_2)^2-M^2)(q-k_1)^2}
\eea

and finally for $\gm_4$
\bea 
(\mu^2)^{2\varepsilon}
\int \dd^d k_2 \frac{1}{k_2^2 (q-k_2)^2}
{\cal T}_{p_1,k_2} \frac{1}{(p_1+k_2)^2-M^2}
\nn \\ \times
\int \dd^d k_1 \frac{1}{k_1^2 (q-k_1)^2}
{\cal T}_{k_1,k_2} \frac{1}{(k_1-k_2)^2-M^2} .
\eea

The results for these three integrals look as follows.
The subgraph $\gm_2$ yields

\be
\frac{1}{(4\pi)^d} \frac{1}{M^4} \left( \frac{\mu^2}{M^2} \right)^{\eps}
\left( -\frac{\mu^2}{q^2} \right)^{\eps}
\sum_{n=0}^{\infty} c^{(2)}_n (\eps) (q^2/M^2)^n ,
\label{C2}
\ee
where
\bea
c^{(2)}_n (\eps) =
\sum_{{i_1,i_2,n_3 \geq 0,\; i_1+i_2+n_3 \; \mbox{\scriptsize even}}
\atop {i_1+i_2+n_3 \leq 2n}} \sum_{j_3 \geq 0}^{(n_3,-2)}
(-1)^{(i_1+i_2+n_3)/2} \;\frac{(n-(i_1+i_2-n_3)/2)!}{(n-(i_1+i_2+n_3)/2)!}
\nn \\
\times \frac{i_1!i_2! \theta (i_1+i_2-j_3)
\theta (i_1-i_2+j_3) \theta (-i_1+i_2+j_3)}{((n_3-j_3)/2)!
((i_1+i_2-j_3)/2)! ((i_1-i_2+j_3)/2)! ((-i_1+i_2+j_3)/2)!}
\nn \\
\times
\frac{\Gm(\eps)\Gm(1-\eps)
\Gm((i_1+i_2-j_3)/2+1-\eps)}{\Gm((i_1+i_2-j_3)/2+2-2\eps)}
\nn \\
\times C(2+i_1 +i_2, 2+n-(i_1+i_2-n_3)/2; (i_1+i_2+n_3)/2),
\eea
and
\be
C(r_1, r_2; s) =
\frac{\Gm(r_1+r_2 -s-\frac{d}{2})\Gm(s-r_2 +\frac{d}{2})}{\Gm(r_1)\Gm(s+\frac{d}{2})} .
\ee
Note that this expression is obtained from the corresponding contribution
of the non-naive part (\ref{nonn}) for our first diagram by the change
\bea
C(2+i_1 +i_2, 2+n-(i_1+i_2-n_3)/2; (i_1+i_2+n_3)/2)
\nn \\ \to C(n+(i_1+i_2+n_3)/2+4; (i_1+i_2+n_3)/2) . \nn
\eea

The contribution from $\gm_3$ takes the following explicit form:
\be
\frac{1}{(4\pi)^d} \frac{1}{M^4} \left( \frac{\mu^2}{M^2} \right)^{\eps}
\left( -\frac{\mu^2}{q^2} \right)^{\eps}
\sum_{n=0}^{\infty} c^{(3)}_n (\eps) (q^2/M^2)^n ,
\label{C3}
\ee
where
\bea
c^{(3)}_n (\eps) =
(-1)^n \sum_{{i_1 \geq 0}
\atop {i_1-(n_1-j_1)/2 \leq n}} \sum_{n_1 = 0}^{i_1}
\sum_{j_1 \geq 0}^{(n_1,-2)}
(-1)^{i_1+j_1} 
\frac{i_1!} {(i_1-n_1)! ((n_1-j_1)/2)!}
\nn \\
\times \frac{\Gm(2-\frac{d}{2})\Gm(\frac{d}{2}-1)
\Gm(n-i_1+(n_1-j_1)/2+\frac{d}{2}-1)}{\Gm(n-i_1+(n_1-j_1)/2+\frac{d}{2})}
\nn \\ \times
C(j_1 +1, i_1+2; (n_1+j_1)/2) .
\eea

Finally, for $\gm_4$ we obtain
\be
\frac{1}{(4\pi)^d} \frac{1}{M^4} \left( -\frac{\mu^2}{q^2} \right)^{2\eps}
\sum_{n=0}^{\infty} c^{(4)}_n (\eps) (q^2/M^2)^n ,
\label{C4}
\ee
with
\be
c^{(4)}_n (\eps) =
(-1)^n (\Gm(2-\frac{d}{2})\Gm(\frac{d}{2}-1))^2 \sum_{j =0}^n
\frac{\Gm(j+\frac{d}{2}-1)\Gm(n-j+\frac{d}{2}-1)}{\Gm(j+d-2)\Gm(n-j+d-2)} .
\ee

After summing up all four contributions we see that the double and
single poles in $\eps$ cancel as well as the scale parameter $\mu$,
with the result
\begin{eqnarray}
F_{\Gamma} (q^2,M^2) =
\frac{1}{M^4} \sum_{n=0}^{\infty} \sum_{j=0}^{2}
f_{jn} (q^2/M^2)^n \ln^j (-q^2/M^2) \nn \\
\label{AE22}
\equiv \frac{1}{M^4} \left\{ f_0 (q^2/M^2 ) + f_1 (q^2/M^2 ) \ln (-q^2/M^2)
+ f_2 (q^2/M^2 ) \ln^2 (-q^2/M^2)\right\} ,
\end{eqnarray}
where the $f_{jn}$ are now obtained in terms of rational numbers and
a $\zeta (2)$ contained in $f_{0n}$.

   The next surprise is that $f_2(x)$ can be summed analytically, yielding\\
$(x=q^2/M^2)$
\begin{eqnarray}
f_2(x)&=&ln^2(1+x)/x^2 ~~~~~~~~~~~~~for~~ x \ge -1, \nonumber \\
      &=&(ln\mid 1+x \mid -i \pi)^2/x^2 ~~~for~~ x < -1.
\end{eqnarray}
Thus, as in $Case$~1, we have to Pad\'{e} approximate $f_0$ and $f_1$ only.
The $\zeta (2)$ term in $f_0$ can in principle also be summed, i.e.
\be
f_0(x)= \tilde{f}_0(x) + 2 \zeta (2) f_2(x),
\ee
with $\tilde{f}_0(x)$ having only rational numbers as expansion coefficients.
This splitting does, however, worsen the convergence properties of the series
for $f_0$ and therefore has no practical meaning. 
Apart from  that the situation
is similar as in $Case$~1, i.e. $f_0$ has 
two thresholds starting at $q^2=\pm M^2$,
respectively
(indeed solving the Landau equations with $m_1=m_2=m_3=m_4=0$ one
obtains the additional pseudothreshold $q^2=-m_5^2$ ). The function
$f_1$, however, is real for timelike $q^2$. In the spacelike region
the cancellation between the imaginary parts 
has now to take place between three
complex functions. This is shown in Table 5, but apparently this cancellation
is not as perfect as in $Case$~1. This is due to a relatively bad convergence
of the Pad\'{e}'s for  $f_0$  (with 30 coefficients) while the convergence
properties of  $f_1$ are much better and  $f_2$ is anyway given analytically.
Fortunately the convergence of $f_0$ is much better in the timelike region
(see Table 6) so
that up to $q^2=10M^2$ the achieved precision is at least 3 decimals. Close to
the second threshold at $q^2=M^2$ the convergence is indeed excellent. It should
be noted that for the physical application we have in mind, i.e. $Z \to b\bar{b}$,
this is just the case of interest. It is worthwhile to note the sharp increase
for low $q^2$, in particular in the timelike region, due to $ln^2(-q^2/M^2)$.

\section{Conclusions}

We have presented a modification of the method of \cite{ft} for the
case when massless thresholds are involved. 
To do this, the starting point was
the asymptotic large mass expansion, rather than the Taylor expansion
in the external momenta.
Using the explicit formulae for the coefficients of this expansion
we obtained a finite sum of powers of the logarithm of the external
momenta times power series in $q^2/M^2$ and then 
applied the technique of conformal
mapping and summation by Pad\'e approximants to each of these
series separately.

We have shown that this new strategy enables us to obtain high precision
numerical values in the domain of physical interest 
including the region beyond the
second threshold in spite of the fact that the initial Feynman
diagrams have their first threshold at the origin of the complex plane.
Below the second threshold, in a certain domain around $q^2=0$,
the functions factorizing powers of $ln(-q^2/M^2)$ can in general be
assumed to be analytic so that even without Pad\'{e} approximants
their power series converge. At the second threshold with definiteness
one can only say that at least one of them must have a cut starting.
This is confirmed by our calculation, i.e. the application of conformal 
mapping and Pad\'{e} approximants (recall the observation that in
$Case$ 5 $f_1$ is real for timelike $q^2$). The occurrence of 
``ghost thresholds'' appears to be related to pseudothresholds
obtained from the Landau equations.

\bigskip
\noindent
{\large{\bf Acknowledgement}}

\medskip
We are grateful to J.B.Tausk for helpful comments concerning
the ``ghost thresholds''. O.T. gratefully acknowledges financial
support by the BMBF, V.S. from the European Community under the
project Human Capital and Mobility.
J.F. thankfully acknowledges financial support by the
DFG for travel expences to Aspen, where part of this work was 
performed and to DESY, Zeuthen to finish it.

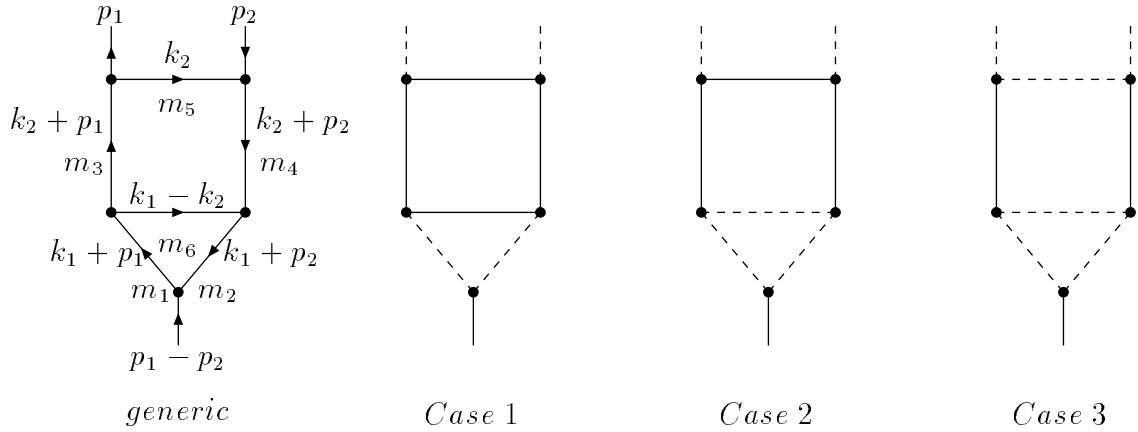
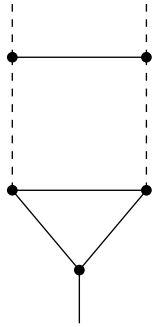
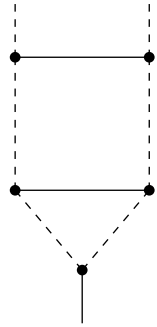
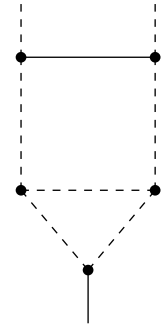
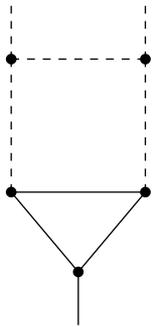
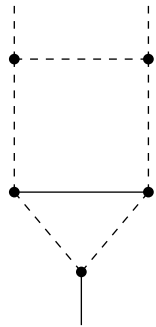
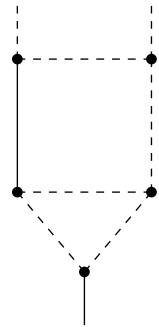
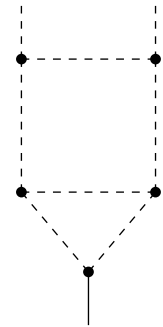
\end{document}